\documentclass[twocolumn,twoside,nofootinbib]{revtex4}
\usepackage{graphicx}
\usepackage{fancyhdr}
\begin{document}

\title{Production of Quarkonia and Heavy Flavor States in ATLAS}

%

\author{Sally Seidel\\
on behalf of the ATLAS Collaboration}
\affiliation{Department of Physics and Astronomy, University of New Mexico, MSC 07 4220, Albuquerque, NM 87131 USA}

\begin{abstract}
  Two recent analyses of data collected with the ATLAS detector at the Large
  Hadron Collider are presented.  In the first, a measurement of $b$ hadron pair production is presented,
  based on a data set corresponding to an integrated luminosity of 11.4~fb$^{-1}$
  of proton-proton collisions recorded at $\sqrt{s}=8$ TeV.  Events are selected
  in which a $b$ hadron is reconstructed in a decay channel containing a muon.
  Results are presented in a fiducial volume defined by kinematic requirements
  on three muons based on those used in the analysis. The fiducial cross section
  is measured to be $17.7 \pm 0.1({\rm stat})\pm 2.0({\rm syst.})$ nb.  A number of
  normalized differential cross sections are also measured and compared to
  predictions from the {\sc Pythia8}, {\sc Herwig++},
  {\sc MadGraph5\_aMC@NLO}+{\sc Pythia8}, and {\sc Sherpa}
  event generators, providing new constraints on heavy flavor production.  In the
  second analysis, the modification of the production of $J/\psi$, $\psi(2S)$,
  and $\Upsilon(nS)~(n=1,2,3)$ in $p+$Pb collisions with respect to their
  production in $pp$ collisions has been studied.  The $p+$Pb and $pp$ data sets
  correspond to integrated luminosities of 28~nb$^{-1}$ and 25~pb$^{-1}$ respectively,
 collected at a center-of-mass energy per nucleon pair of 5.02 TeV. The
  quarkonium states are reconstructed in the dimuon decay channel.  The yields
  of $J/\psi$ and $\psi(2S)$ are separated into prompt and non-prompt sources.
  The measured quarkonium differential cross sections are presented as a function
  of rapidity and transverse momentum, as is the nuclear modification factor
  $R_{p{\rm Pb}}$ for the $J/\psi$ and $\Upsilon(nS)$.  No significant modification
  of the $J/\psi$ production is observed, while $\Upsilon(nS)$ production is
  found to be suppressed at low transverse momentum in $p+$Pb collisions
  relative to $pp$ collisions.  The production of excited charmonium and
  bottomonium states is found to be suppressed relative to that of the ground
  states in central $p+$Pb collisions.
  
\end{abstract}

\maketitle

\thispagestyle{fancy}


\section{Introduction}
Two recent measurements involving quarkonium production and $b$- and $c$-quarks
(hereafter called heavy flavor states), by the ATLAS Experiment~\cite{atlas} at the Large Hadron Collider, are presented.\footnote{Copyright 2019 CERN for
the benefit of the ATLAS Collaboration CC-BY-4.0 license.} 

\section{The b-Hadron Pair Production Cross Section}
The production of heavy flavor states in proton-proton collisions
provides a fruitful testing ground for the predictions of quantum chromodynamics (QCD).  The mass of
the $b$-quark introduces a scale, and the factorization of QCD calculations into parton distribution
functions, hard matrix element, and softer parton shower components allows the mass to be introduced
at different stages.  Furthermore there are several possible schemes for the inclusion of the heavy
quark masses at the various stages.  The optimal settings must be determined by comparison to data.
Previous measurements have highlighted disagreements both between different theoretical predictions
and between those predictions and the data.  The realm of small angle $b {\overline b}$ pair production
is particularly sensitive to the details of the calculations but remains only loosely constrained by
data.  Searches for Higgs bosons produced in association with a vector boson and decaying to a
$b {\overline b}$ pair rely extensively on the modelling of the background arising from QCD production
of $b {\overline b}$ pairs in association with vector bosons.

Motivated by these facts, a measurement~\cite{bhadron} has been made of the production of two $b$-hadrons, where one
$b$-hadron decays to $J/\psi(\rightarrow \mu^+ \mu^-) +X$ and the other to $\mu +Y$, resulting in three
muons in the final state.  The signal definition includes $J/\psi$ mesons produced via feed-down
from excited charmonium states as well as muons produced in semileptonic cascade decays.  To probe
$b$-hadron production, several differential cross sections are measured, based on the kinematics
of the $J/\psi$ (reconstructed from two muons) and the third muon.  The variables considered are:
\begin{itemize}
\item the azimuthal separation $\Delta \phi(J/\psi,\mu)$ between the $J/\psi$ and the third muon;
\item the transverse momentum of the three-muon system, $p_{\rm T}(J/\psi,\mu)$;
\item the separation between the $J/\psi$ and the third muon in the azimuth-rapidity plane, $\Delta R(J/\psi,\mu)$;
\item the separation in rapidity  $\Delta y(J/\psi,\mu)$ between the $J/\psi$ and the third muon;
\item the magnitude of the average rapidity of the $J/\psi$ and the third muon, $y_{\rm boost}$;
\item the mass of the three-muon system, $m(J/\psi,\mu)$;
\item the ratio of the transverse momentum of the three-muon system to the invariant mass of the
  three-muon system, $p_{\rm T}/m$, and its inverse, $m/p_{\rm T}$.
\end{itemize}
The differential cross sections are measured and compared to predictions from several Monte Carlo generators.

Events are selected using a dimuon trigger where the muons are required to have opposite charge, be consistent
with originating at the same production vertex, have $p_{\rm T}(\mu) > 4$ GeV and pseudorapidity $|\eta(\mu)| < 2.4$, and satisfy
$2.5 <  m(\mu^+\mu^-) < 4.3$ GeV. The integrated luminosity of the data set is 11.4 fb$^{-1}$.  Following
selection of the primary vertex (defined as the vertex formed from at least two tracks, each with $p_{\rm T} > 400$ MeV,
that has the largest summed track $p_{\rm T}^2$ in the event), muon candidates are formed, and $J/\psi$ candidates are
reconstructed from them.
The third muon is then selected as the highest-$p_{\rm T}$ muon not used in the $J/\psi$ candidate reconstruction.
To extract the $J/\psi$ mesons resulting from the decay of a $b$-hadron, a simultaneous fit is performed upon
the distributions of dimuon mass and the pseudo-proper decay time $\tau$, defined as
$$\tau=L_{xy}\cdot m(J/\psi_{\rm PDG})/p_{\rm T}(\mu^+\mu^-),$$
where $L_{xy}$ is the signed transverse distance between the primary vertex and the dimuon vertex:
$$L_{xy}={\vec{L}\cdot\vec{p}_T \over p_{\rm T}}.$$  Because
$J/\psi$ mesons produced by most $b$-hadron decays are non-prompt, a selection $\tau > 0.25$ mm/$c$ is applied.

Following the addition of the third muon, the yield of signal events relative to background is further
improved using a simultaneous fit to the transverse impact parameter significance and the output of a boosted
decision tree trained to separate signal muons from misreconstructed muons.  The transverse impact parameter
significance is defined as $S_{d_0} \equiv d_0/\sigma_{d_0},$ where $d_0$ is the signed distance of closest approach of the
track to the primary vertex point in the $r-\phi$ projection and $\sigma_{d_0}$ is the unsigned uncertainty
on the reconstructed $d_0$.  Some remaining irreducible sources of background are then subtracted from the fitted yields.
Corrections for the effects of the selection on $\tau$ and for the effects of detector resolution are applied, and the
analysis is repeated for every kinematic bin for each differential cross section, resulting in the cross section in
that bin.

The total measured cross section in the fiducial region~\cite{bhadron} is
$\sigma(B(\rightarrow J/\psi[\rightarrow \mu^+ \mu^-]+X)B(\rightarrow \mu +X))=17.7 \pm 0.1 (\rm stat) \pm 2.0 (\rm syst) {\rm nb}.$

The data are compared to predictions by various simulations including that for inclusive $b$-hadron pairs taken from
{\sc Pythia8.186}.  These explore several options for the $g\rightarrow b{\overline b}$ splitting kernel as this process
dominates small-angle $b$-hadron production.  The settings explore whether to use the relative $p_{\rm T}$ or mass of the
splitting to set the scale when determining the value of $\alpha_s$ to use in that splitting.  It is found that
in general, {\sc Pythia8} does not reproduce the shape of the angular distributions in data within uncertainties, and
the $p_{\rm T}$-based scale splitting kernels generally give a better description of the low $\Delta R(J/\psi,\mu)$ region.

\begin{figure}[hbtp]
  \centering
  \includegraphics[width=3in]{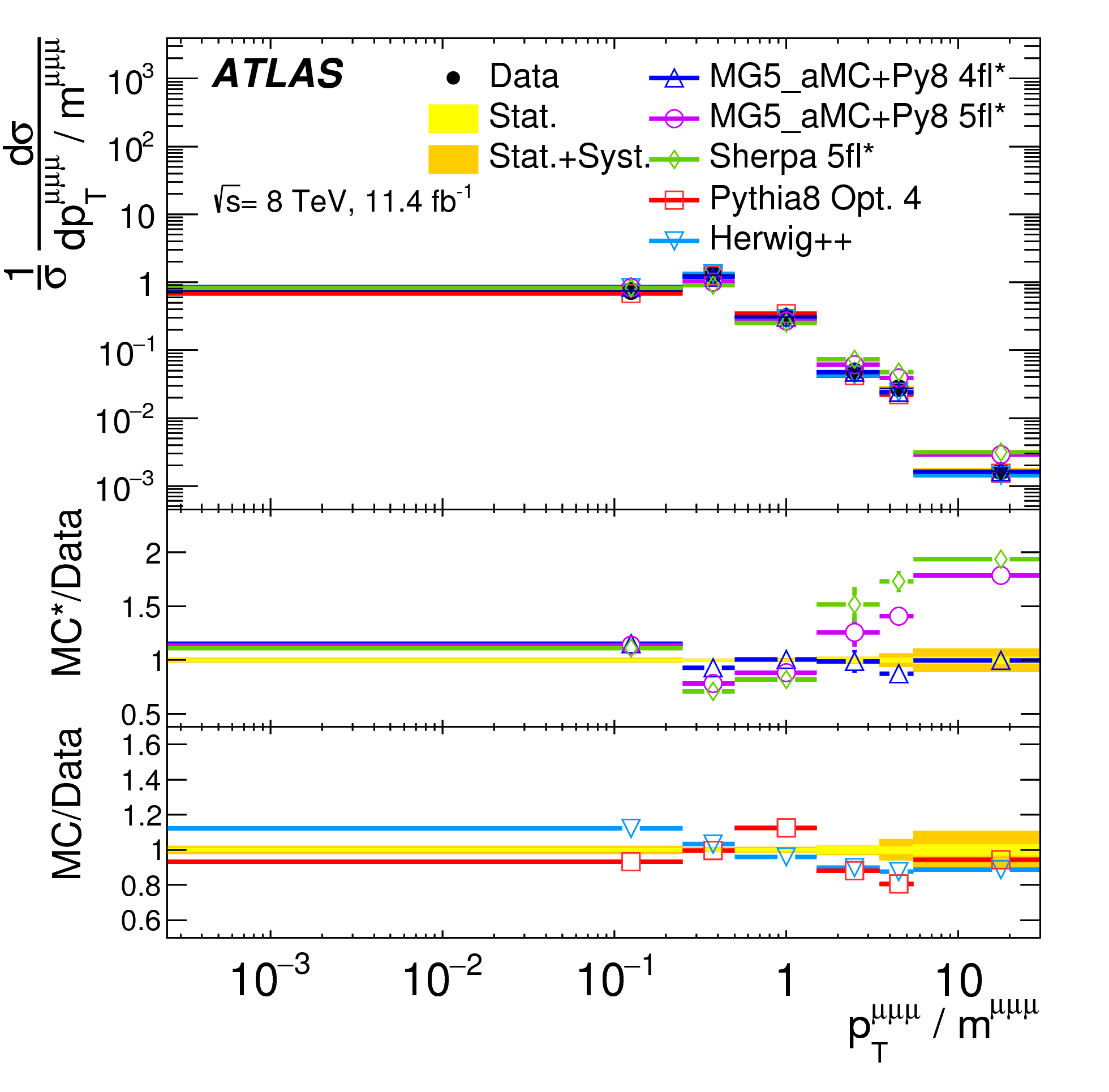}
  \caption{The measured normalized differential cross section as a function of $p_{\rm T}/m$~\cite{bhadron}.  Comparisons are made with predictions
    of {\sc Pythia8} and {\sc Herwig++}.  {\sc MadGraph5\_aMC@NLO+Pythia8} and {\sc Sherpa} predictions are also compared having
    been corrected from the two-$b$-hadron production to the three-muon final state via transfer functions (indicated with $^*$).
    The {\sc Pythia8} ``Opt. 4'' gluon splitting parameter settings use a splitting kernel $z^2+(1-z)^2+8r_qz(1-z)$, normalized
    so that the $z$-integrated rate is $(\beta/3)(1+r/2)$, and with an additional suppression factor
    $(1-m_{qq}^2/m^2_{\rm dipole})^3$, which reduces the rate of high-mass $q{\overline q}$ pairs.}
  \label{bhadron-figure}
\end{figure}

The comparisons are then extended to predictions using {\sc Herwig++}, {\sc MadGraph5\_aMC@NLOv2.2.2} interfaced to the {\sc Pythia8.186}
parton shower model, and {\sc Sherpa2.1.1}.  Options with 5-flavor and 4-flavor schemes are compared.  The comparisons
indicate that (1) agreement with data is slightly better for {\sc Herwig++} than {\sc Pythia8} in the $\Delta R(J/\psi,\mu)$
distribution; (2) the 4-flavor prediction by {\sc MadGraph+Pythia8} is generally closer in shape to the data than the
5-flavor prediction; (3) in $\Delta y(J/\psi,\mu)$, the {\sc MadGraph+Pythia} and {\sc Sherpa} predictions provide a good description
of the data; (4) in $y_{\rm boost}$, all models are comparable; (5) the 5-flavor {\sc MadGraph+Pythia8} prediction lies
closer to data than the 4-flavor one in the low mass distribution; and (6) at high values of $p_{\rm T}/m$, the 4-flavor
prediction describes the data better than the 5-flavor prediction.  Considering all distributions, the 4-flavor
{\sc MadGraph+Pythia8} prediction provides the best description of the data overall, the predictions of {\sc Pythia8} and {\sc Herwig++}
are comparable, and of the {\sc Pythia8} options considered, the $p_{\rm T}$-based splitting kernel gives the best agreement with data. Figure~\ref{bhadron-figure} compares the various predictions to the measured normalized differential cross section in $p_{\rm T}/m$.

\section{Quarkonium Production in Proton-Lead and Proton-Proton Collisions}

In order to understand quarkonium
yields in nucleus-nucleus ($A+A$) collisions, it is necessary to disentangle effects
due to interaction between quarkonium and the quark-gluon plasma (QGP) medium
from those that can be ascribed to cold nuclear matter (CNM).
In proton (deuteron)-nucleus collisions, $p(d)+A$, the formation
of a large region of deconfined and hot QGP matter is
not expected to occur. Therefore,  suppression of
quarkonium yields in these systems with respect to $pp$ collisions
has traditionally been attributed to CNM effects.

Among the CNM effects, three primary initial-state
effects include: modifications of the nuclear parton distribution
functions, parton saturation effects in the incident
nucleus, and parton energy loss through interactions
with the nuclear medium. On the other hand,
the absorption of the heavy quark-antiquark pair through
interactions with the co-moving nuclear medium is
considered to be a final-state effect. In proton-lead ($p+$Pb)
collisions, the modification of quarkonium production with
respect to that in $pp$ collisions may be quantified by the
nuclear modification factor, $R_{p{\rm Pb}}$, which is defined as the
ratio of the quarkonium production cross section in $p+$Pb
collisions to the cross section measured in $pp$ collisions at
the same center-of-mass energy, scaled by the number of
nucleons in the lead nucleus:

$$R_{p{\rm Pb}}={1 \over 208} {\sigma^{O(nS)}_{p+{\rm Pb}} \over \sigma^{O(nS)}_{pp}},$$

\noindent where $O(nS)$ represents one of five measured quarkonium
states: $J/\psi$, $\psi(2S)$, $\Upsilon(1S)$, $\Upsilon(2S)$, and $\Upsilon(3S)$. 

The CNM effects in excited quarkonium states with respect
to the ground state can be quantified by the double ratio,
defined as:

$$\rho^{O(nS)/O(1S)}_{p{\rm Pb}}={R_{p{\rm Pb}}(O(nS)) \over R_{p{\rm Pb}}(O(1S))}$$

\noindent where $n = 2$ for charmonium and $n = 2$ or 3 for bottomonium.
In the double ratio, most sources of detector systematic
uncertainty cancel, and measurements of this quantity by
different experiments can easily be compared. The initial state
effects are expected to be largely cancelled out in the double
ratio due to the same modifications affecting partons before
the formation of both quarkonium states, so measuring the relative
suppression of different quarkonium states should help
in understanding the properties of the final-state effects separately
from the initial ones.

Several classes of experimental measurements
are reported~\cite{heavyion}: differential
production cross sections of $J/\psi$, $\psi(2S)$, and
$\Upsilon(nS)~(n = 1, 2, 3)$ in $pp$ collisions at
$\sqrt{s} = 5.02$ TeV and $p+$Pb
collisions at $\sqrt{s_{NN}}= 5.02$ TeV; center-of-mass
rapidity dependence and transverse momentum dependence
of $J/\psi$ and $\Upsilon(1S)$ nuclear modification factors, $R_{p{\rm Pb}}$;
and charmonium and bottomonium double ratios, $\rho^{O(nS)/O(1S)}_{p{\rm Pb}}$.

The $p+$Pb collisions in this analysis result from the interactions of a proton
beam with an energy of 4 TeV and a lead beam with an
energy of 1.58 TeV per nucleon.  In the $p+$Pb collision configuration, the
proton-nucleon center-of-mass rapidity, $y^*$, had a shift of $\Delta y = 0.465$ with respect to $y$ in the laboratory frame. After
60\% of the data were recorded, the directions of the proton
and lead beams were reversed.  All data from
both periods are presented in $y^*$. The proton beam always travels in the direction of
positive $y^∗$.

Candidate events in $p+$Pb collisions were collected with a
dimuon trigger.  Pairs of muon candidates satisfying various quality
requirements, and with opposite charges, are selected as
quarkonia candidate pairs.  In order to characterize the $p+$Pb collision geometry, each
event is assigned to a centrality class based on the total
transverse energy measured in the Forward Calorimeter on the Pb-going
side (backwards). Collisions with more (fewer) participating
nucleons are referred to as central (peripheral).

The double-differential cross section multiplied by the
dimuon decay branching fraction is calculated for each measurement
interval as:

$${d^2 \sigma_{O(nS)} \over dp_{\rm T}dy^*} \times B(O(nS) \rightarrow \mu^+\mu^-) = {N_{O(nS)} \over {\Delta p_{\rm T} \times \Delta y \times L}},$$ 

\noindent where $L$ is the integrated luminosity, $\Delta p_{\rm T}$ and $\Delta y$ are
the interval sizes in dimuon transverse momentum
and center-of-mass rapidity, respectively, and $N_{O(nS)}$
is the observed quarkonium yield in the kinematic interval
under study, extracted from fits and corrected for acceptance,
trigger and reconstruction efficiencies.

The charmonium yield determination decomposes the yields
into two sources of muon pairs: prompt and
non-prompt. The prompt $J/\psi$ and $\psi(2S)$ signal originates
from the strong production of short-lived particles, including
feed-down from other short-lived charmonium states, while
non-prompt refers to $J/\psi$ and $\psi(2S)$ mesons which are the
decay products of $b$-hadrons. To distinguish between these
prompt and non-prompt processes, the pseudo-proper decay time, $\tau_{\mu\mu}$,
is used.
Two-dimensional unbinned
maximum-likelihood fits are performed on weighted distributions of
the dimuon invariant mass ($m_{\mu\mu}$) and pseudo-proper decay time
to extract prompt and non-prompt signal yields,
in intervals of $p_{\rm T}^{\mu\mu}$,
rapidity, and centrality.  To obtain the acceptance corrections, $J/\psi$
acceptance is applied to events with $m_{\mu\mu} < 3.2$ GeV, $\psi(2S)$
acceptance is applied to events with $m_{\mu\mu} > 3.5$ GeV, and a
linear interpolation of the two acceptances is used for events
with $3.2 < m_{\mu\mu} < 3.5$ GeV.

The yields of bottomonium states are obtained by performing
unbinned maximum likelihood fits of the weighted invariant
mass distribution, in intervals of $p_{\rm T}^{\mu\mu}$, rapidity, and centrality.
Due to overlaps between the invariant mass peaks of different
bottomonium states, the linear acceptance interpolation
used for the charmonium states is not appropriate to the bottomonium
states. Instead, each interval is fitted three times
to extract the corrected yields of the three different bottomonium
states.

\begin{figure}[h]
\centering
\includegraphics[width=80mm]{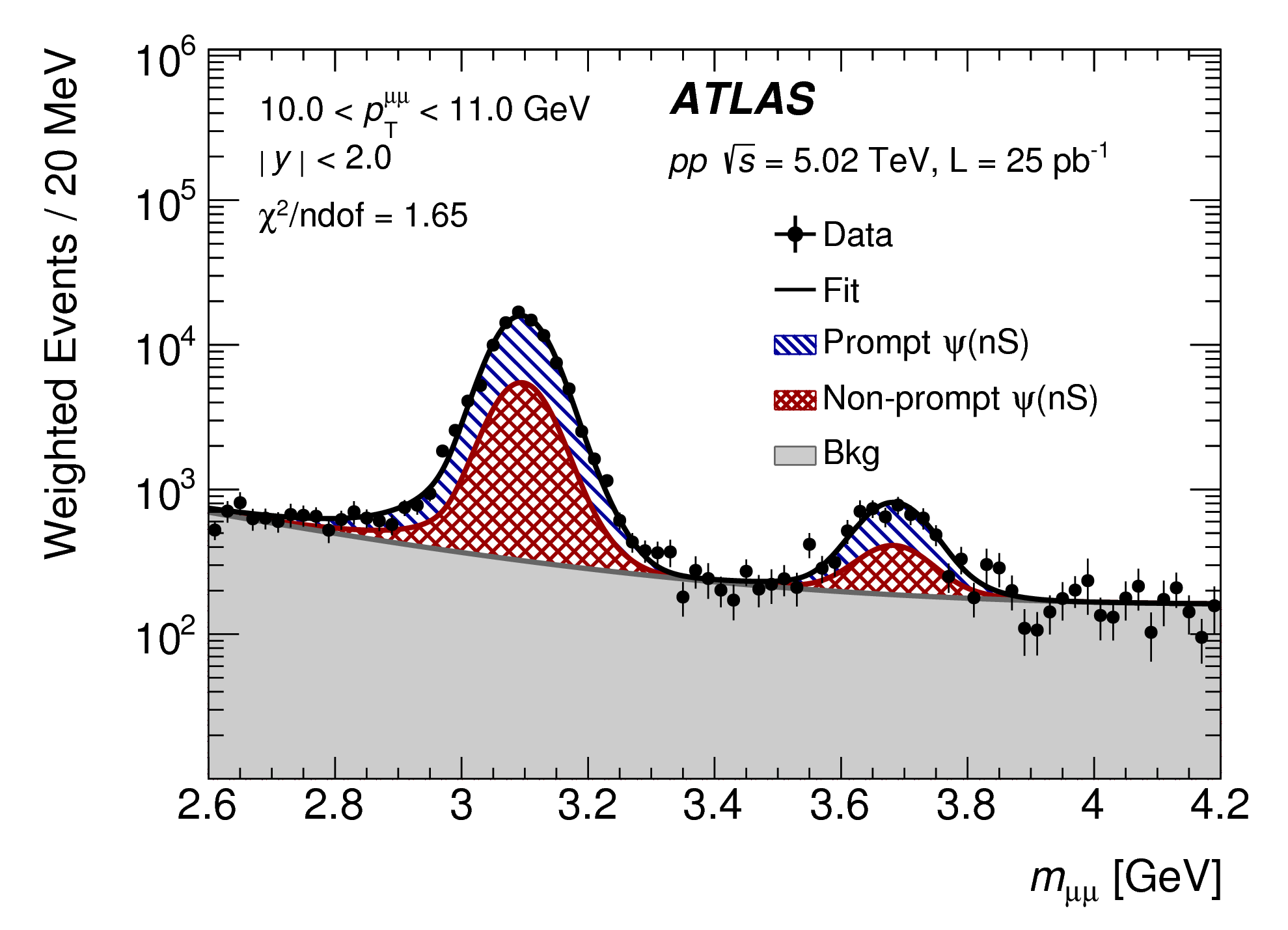}
\includegraphics[width=80mm]{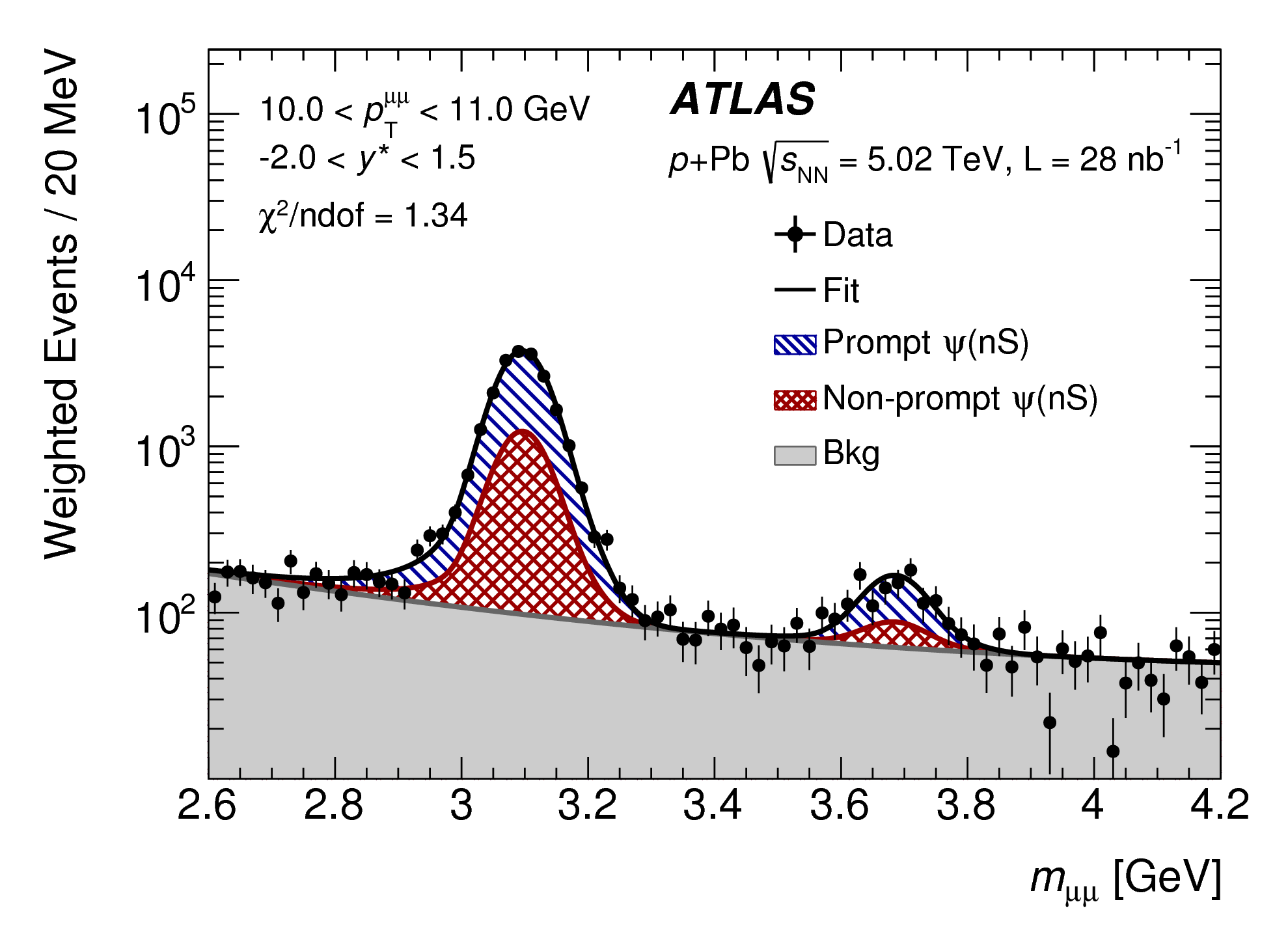}
\caption{Projection of the charmonium fit results~\cite{heavyion} onto dimuon invariant
  mass $m_{\mu\mu}$ for $pp$ collisions at $\sqrt{s}=5.02$ TeV for the kinematic ranges
  $10<p_{\rm T}^{\mu\mu}<11$ GeV and $|y|<2.0$ (upper), and $p+$Pb collisions
  at $\sqrt{s_{NN}}=5.02$ TeV for the kinematic ranges $10<p_{\rm T}^{\mu\mu}<11$ GeV and $-2.0<y^*<1.5$ (lower).}
\label{heavyion-fig1}
\end{figure}

Figure~\ref{heavyion-fig1} shows examples of charmonium fit projections
onto the invariant mass axis.

Following the yield correction and signal extraction, the cross
sections of the five quarkonium states are
measured differentially in transverse momentum and rapidity. The results for non-prompt $J/\psi$ and
$\psi(2S)$ cross sections
in $pp$ collisions at $\sqrt{s} = 5.02$ TeV are compared to fixed order
next-to-leading-logarithm (FONLL) predictions. The FONLL uncertainties include renormalization
and factorization scale variations, charm quark mass,
and parton distribution functions uncertainties. The measured non-prompt charmonium production
cross sections agree with the FONLL predictions within
uncertainties over the measured $p_{\rm T}$ range.

The measured prompt $J/\psi$ and $\psi(2S)$ cross sections are compared with non-relativistic
QCD (NRQCD) predictions. The theory predictions are
based on long-distance matrix elements (LDMEs) with uncertainties originating from the choice
of scale, charm quark mass, and LDMEs. The production cross section
of $\Upsilon(nS)$ in $pp$ collisions is  compared to similar NRQCD
model calculations.

The results for prompt and non-prompt production cross
sections of $J/\psi$ and $\psi(2S)$ in $p+$Pb collisions at $\sqrt{s_{NN}}=5.02$ TeV
are consistent with previous results within uncertainties.
The measured differential production cross sections of
$\Upsilon(2S)$ and $\Upsilon(3S)$ in $p+$Pb collisions are combined to obtain
stable rapidity dependence of the production cross section.

Both the prompt and
non-prompt $J/\psi$ $R_{p{\rm Pb}}$ are consistent with unity across the
$p_{\rm T}$ range from 8 to 40 GeV. No significant rapidity dependence is observed.

Using $R_{p{\rm Pb}}$,
the $\Upsilon(1S)$ production in $p+$Pb collisions is found to be
suppressed compared to $pp$ collisions at $p_{\rm T} <15$ GeV,
and it increases with $p_{\rm T}$. Low $p_{\rm T}$ $\Upsilon(1S)$ states can probe a
smaller Bjorken-$x$ region than can $J/\psi$ measured in
$8 < p_{\rm T} < 40$ GeV, so the observed suppression of
$\Upsilon(1S)$ production at low $p_{\rm T}$ may come from the reduction
of hard-scattering cross sections due to stronger nuclear PDF
shadowing at smaller Bjorken-$x$. No significant rapidity
dependence is observed.

The prompt charmonium double ratio is found by ATLAS to decrease
slightly from the backward to the forward center-of-mass
rapidity. The prompt $\psi(2S)$ production is suppressed with
respect to prompt $J/\psi$ production in $p+$Pb collisions with
a significance of one standard deviation. The production of
excited bottomonium states, $\Upsilon(2S)$ and $\Upsilon(3S)$, is found to
be suppressed at the LHC with respect to $\Upsilon(1S)$ in the integrated kinematic
ranges of $p_{\rm T} < 40$ GeV and $ -2 < y^* < 1.5$ in $p+$Pb
collisions with significance at the level of two standard deviations.
Both the prompt $\psi(2S)$ to $J/\psi$ and $\Upsilon(2S)$ to $\Upsilon(1S)$
double ratios show decreasing behavior in more central collisions.
The decreasing trends from peripheral to central are
at the significance level of one standard deviation. Thus a stronger
cold nuclear matter effect is observed in excited quarkonium
states compared to that in ground states.
This work expands the kinematic range of measured charmonium
and bottomonium cross sections in $pp$ and $p+$Pb
collisions. It thus serves as an additional dataset for constraining
different models of cold nuclear matter effects and
quantifying heavy quarkonium production.

\begin{acknowledgments}
This work was supported by U.S. Department of Energy grant DE-SC0019101.
\end{acknowledgments}

\bigskip 

\begin{thebibliography}{9}   

\bibitem{atlas} ATLAS Collaboration, {\it The ATLAS experiment at the CERN Large Hadron Collider,} 2008 JINST {\bf 3} S08003.  
  
\bibitem{bhadron} ATLAS Collaboration, {\it Measurement of b-hadron pair production with the ATLAS detector in proton-proton collisions at $\sqrt{s}=8$ TeV,} JHEP 11 (2017) 062.

\bibitem{heavyion} ATLAS Collaboration, {\it Measurement of quarkonium production in proton-lead and proton-proton collisions at 5.02 TeV with the ATLAS detector,} Eur. Phys. J. C (2018) 78:171.


\end{thebibliography}

\end{document}